\documentclass[proceedings]{JHEP3}
\usepackage{amsfonts}

\usepackage{amsmath}
\usepackage{epsfig,multicol}

\newbox\mybox

\newcommand\fverb{\setbox\mybox=\hbox\bgroup\verb}
\newcommand\fverbdo{\egroup\medskip\noindent\fbox{\unhbox\mybox}\ }
\newcommand\fverbit{\egroup\item[\fbox{\unhbox\mybox}]}
\conference{$\mathcal{PT}$-symmetric extensions of the  supersymmetric Korteweg-de Vries 
Equation}
\abstract{We discuss several $\mathcal{PT}$-symmetric deformations of superderivatives. 
Based on these various possibilities, we propose new families of complex $\mathcal{PT}$-symmetric deformations of the supersymmetric Korteweg-de Vries equation. Some of these new models are mere fermionic extensions of the former in the sense that they
are formulated in terms of superspace valued superfields containing bosonic and fermionic 
fields, breaking however the supersymmetry invariance. Nonetheless, we also find 
extensions, which may be viewed as new supersymmetric Korteweg-de Vries equation. 
Moreover, we show that these deformations allow for a non-Hermitian Hamiltonian 
formulation and construct three charges associated to the corresponding flow.}

\title{$\mathcal{PT}$-symmetric extensions of the supersymmetric Korteweg-de Vries
equation}
\author{Bijan Bagchi$^\circ$ and Andreas Fring$^\bullet$ \\
$\circ$ Department of Applied Mathematics, University of Calcutta \\
\quad 92 Acharya Prafulla Chandra Road, Kolkata 700 009, India\\
$\bullet$ Centre for Mathematical Science, City University\\
\quad Northampton Square, London EC1V 0HB, UK\\
E-mail: \email{BBagchi123@reddiffmail.com, A.Fring@city.ac.uk}}

\input{tcilatex}

\begin{document}

\section{Introduction}

$\mathcal{PT}$-symmetry, that is the invariance under a simultaneous parity
transformation $\mathcal{P}:x\rightarrow -x$ and time reversal $\mathcal{T}%
:t\rightarrow -t$, is a very desirable property to have in a physical model
without dissipation. For a Hamiltonian system it can be exploited to
guarantee the reality of the corresponding spectrum, even though the
Hamiltonian might be non-Hermitian \cite{EW,Bender:1998ke,Bender:2002vv}.
However, even for non-Hamiltonian systems this principle can be used to
construct interesting new complex extended models, e.g. \cite
{Bender:2006tz,Ali23,Bender:2007gm,Josh,BBCF,AFKdV}. See \cite
{Benderrev,special2} for a review and some recent results of this field of
research.

Here we commence with an integrable model, which are well known to exhibit
many extremely interesting features on the classical as well as on the
quantum level. Due to their rich structure it is a very natural and common
procedure to take these models as starting points and study new models
closely related to them. We intend here to perturb or deform such a model in
a $\mathcal{PT}$-symmetric manner. Concerning integrable models only few
extensions of such type have been constructed. So far several extensions
related to Calogero-Moser-Sutherland models \cite
{Basu-Mallick:2000af,Basu-Mallick:2001ce,Milos,AF,AFActa,AFZ} and the
Korteweg-de Vries (KdV) equations \cite{BBCF,AFKdV} have been investigated.
Based on the observation that also the supersymmetric version of the
KdV-equation (sKdV) is $\mathcal{PT}$-symmetric, the main aim in this
manuscript is to extend these type of analysis to this equation.

Our manuscript is organized as follows: In section 2 we recall some basic
facts about the sKdV-equation and demonstrate how the $\mathcal{PT}$%
-symmetry manifests itself in these equations. We exploit these observations
to discuss various versions of $\mathcal{PT}$-symmetrically deformed
superderivatives and demonstrate how they can be employed to construct new
models. In section 3 we provide a supersymmetric Hamiltonian version of a
such extensions. We state our conclusions in section 4.

\section{$\mathcal{PT}$-symmetric extensions the sKdV equation}

Let us first fix our notations and recall some known facts about the
sKdV-equation. There exist various fermionic extensions of the KdV-equation
in terms of superfields, which are either supersymmetric \cite{Mat} or break
this symmetry \cite{Kuper,Manin} and are therefore mere fermionic
extensions. We take as a starting point the former case and focus on the
one-parameter family of the sKdV-equation as derived first by Mathieu in 
\cite{Mat} 
\begin{equation}
\Phi _{t}=-D^{6}\Phi +\lambda D^{2}(\Phi D\Phi )+(6-2\lambda )D\Phi
D^{2}\Phi .  \label{1}
\end{equation}
Here $\lambda $ is a real constant and $\Phi (x,\theta )$ denotes a
fermionic superfield 
\begin{equation}
\Phi (x,\theta )=\xi (x)+\theta u(x)  \label{22}
\end{equation}
defined in terms of the fermionic (anticommuting) field $\xi (x)$, the usual
bosonic (commuting) KdV field $u(x)$ and the anticommuting superspace
variable $\theta $. Furthermore $D$ in (\ref{1}) denotes the superderivative
defined as
\begin{equation}
D=\theta \partial _{x}+\partial _{\theta }.  \label{D}
\end{equation}
Expanding the superfield $\Phi $ in terms of component fields, as specified
in (\ref{22}), equation (\ref{1}) may be re-written as a set of two coupled
equations
\begin{eqnarray}
u_{t} &=&-u_{xxx}+6uu_{x}-\lambda \xi \xi _{xx},\text{\qquad }  \label{2} \\
\xi _{t} &=&-\xi _{xxx}+(6-\lambda )\xi _{x}u+\lambda \xi u_{x}.  \label{3}
\end{eqnarray}
When $\lambda \rightarrow 0$ or $\xi \rightarrow 0$ equation (\ref{2})
reduces to the standard KdV equation. In superspace the supersymmetry
transformation is realized as
\begin{equation}
\mathcal{SUSY}:x\rightarrow x-\eta \theta ,\quad \theta \rightarrow \theta
+\eta ,  \label{SUSY1}
\end{equation}
with $\eta $ being an anticommuting constant. As a consequence the
superfield and its components transform as
\begin{equation}
\mathcal{SUSY}:\Phi \rightarrow \Phi +\eta u+\theta \eta \xi _{x},\quad
u\rightarrow u+\eta \xi _{x},\quad \xi \rightarrow \xi +\eta u,  \label{SUSY}
\end{equation}
i.e. a bosonic field is related to a fermionic one and vice versa. Equations
(\ref{1}), (\ref{2}) and (\ref{3}) are designed to remain invariant under
the changes (\ref{SUSY}).

In order to see how one can deform the sKdV-equation in a $\mathcal{PT}$%
-symmetric manner, we need to establish first how this symmetry manifests
itself. We observe that the equation (\ref{1}) remains invariant under the
following anti-linear symmetry transformation
\begin{equation}
\mathcal{PT}:t\rightarrow -t,x\rightarrow -x,i\rightarrow -i,\Phi
\rightarrow i\Phi ,D\rightarrow -iD.  \label{spt1}
\end{equation}
As a result of these properties of the superfield and superderivative we
deduce that the component fields and the superspace variable transform as
\begin{equation}
\mathcal{PT}:u\rightarrow u,\xi \rightarrow i\xi ,\theta \rightarrow i\theta
.  \label{spt2}
\end{equation}
These transformations leave the equations (\ref{2}) and (\ref{3}) invariant.
Notice that the $\mathcal{PT}$-transformation is an automorphism and we
still have $\mathcal{PT}^{2}=1$, as it should be.

Before we embark on the task of seeking $\mathcal{PT}$-symmetric extensions
of equation (\ref{1}) or its equivalent component version (\ref{2}), (\ref{3}%
), we shall define some deformations of derivatives and their supersymmetric
counterparts in a more generic fashion.

\subsection{Deformed (super) derivatives}

In the spirit of the construction in \cite{BBCF,AFKdV} we will define some
new superderivatives, which respect the $\mathcal{PT}$-transformation
properties (\ref{spt1}). For this purpose we recall how to employ an
ordinary deformed derivative $\partial _{x,\varepsilon }$ acting on some
arbitrary $\mathcal{PT}$-invariant function $f(x)$ 
\begin{equation}
\partial _{x,\varepsilon }f(x)=-i(if_{x})^{\varepsilon }\text{\qquad \qquad
with \ }\varepsilon \in \mathbb{R}.  \label{a}
\end{equation}
The case $\varepsilon =1$ corresponds to the standard undeformed case.
Notice further that this deformed differential operator acts not
distributively. We define higher derivatives by acting successively with
ordinary derivatives on $\partial _{x,\varepsilon }$ as
\begin{equation}
\partial _{x,\varepsilon }^{n}:=\partial _{x}^{n-1}\partial _{x,\varepsilon
}.  \label{cc}
\end{equation}
Alternatively we could have introduced a nested version of (\ref{a}) or
possibly a mix of $\partial _{x,\varepsilon }$ and $\partial _{x}$ in
succession such as $\partial _{x,\varepsilon }(\partial _{x,\varepsilon
}\ldots (\partial _{x,\varepsilon }f(x)\ldots ))$ or $\partial
_{x,\varepsilon }(\partial _{x}\ldots (\partial _{x,\varepsilon }f(x)\ldots
))$. These latter possibilities do of course also not break the $\mathcal{PT}
$-symmetry, but they would insinuate a much higher degree of non-linearity
than the definition (\ref{cc}). More explicitly the first expressions for (%
\ref{cc}) read
\begin{eqnarray}
\partial _{x,\varepsilon }^{2}f &=&-i\varepsilon (if_{x})^{\varepsilon }%
\frac{f_{xx}}{f_{x}},\quad ~~  \label{d1} \\
\partial _{x,\varepsilon }^{3}f &=&-i\varepsilon (if_{x})^{\varepsilon } 
\left[ \frac{f_{xxx}}{f_{x}}+(\varepsilon -1)\left( \frac{f_{xx}}{f_{x}}%
\right) ^{2}\right] ,  \label{d2} \\
\partial _{x,\varepsilon }^{4}f &=&-i\varepsilon (if_{x})^{\varepsilon } 
\left[ (2+\varepsilon (\varepsilon -3))\left( \frac{f_{xx}}{f_{x}}\right)
^{3}+3(\varepsilon -1)\left( \frac{f_{xx}}{f_{x}}\right) ^{2}f_{xxx}+\frac{%
f_{xxxx}}{f_{x}}\right] .  \label{d3} \\
&&\vdots  \notag
\end{eqnarray}
Note that for $\varepsilon =-1/2$ the bracket in (\ref{d2}) simply becomes a
Schwarzian derivative.

Obviously by construction the derivatives $\partial _{x,\varepsilon }^{n}$
and $\partial _{x,\varepsilon =1}^{n}=\partial _{x}^{n}$ transform in the
same way under a $\mathcal{PT}$-transformation, i.e. $\mathcal{PT}:$ $%
\partial _{x}^{n}\rightarrow (-1)^{n}\partial _{x}^{n}$ and $\mathcal{PT}:$ $%
\partial _{x,\varepsilon }^{n}\rightarrow (-1)^{n}\partial _{x,\varepsilon
}^{n}$, which gives rise to the simple construction principle: In a defining
equation of a particular model replace $\partial _{x}^{n}$ by $\partial
_{x,\varepsilon }^{n}$ in order to introduce a new family of models.

Next we employ these deformations of ordinary derivatives to define a
deformed version of the superderivative (\ref{D}) 
\begin{equation}
D_{\varepsilon }:=\theta \partial _{x,\varepsilon }+\partial _{\theta }.
\end{equation}
Clearly $D$ and $D_{\varepsilon }$ have the same transformation properties
with regard to (\ref{spt1}) and (\ref{spt2}). The derivative with respect to
the superspace variable is left undeformed as there is no natural deformed
counterpart to this. In the deformation of the standard derivative we could
achieve that the minus sign results from the anti-linear nature of the $%
\mathcal{PT}$-operator through the newly introduced factor $i$ rather than
from $\partial _{x}$. In contrast, for the derivative $\partial _{\theta }$
we can not implement this feature, since in that case we have $\mathcal{PT}$%
: $\partial _{\theta }\Phi \rightarrow \partial _{\theta }\Phi $. Depending
now on the way the higher derivatives are defined one may obtain
deformations only acting on the bosonic, fermionic or possibly on both type
of fields. Let us explore these possibilities.

\subsubsection{PT-symmetric superderivatives of bosonic-fermionic type}

As a first option we define higher deformed superderivatives as
\begin{eqnarray}
D_{\varepsilon }^{2} &:&=D_{\varepsilon }D_{\varepsilon }, \\
D_{\varepsilon }^{n} &:&=D^{n-2}D_{\varepsilon }^{2}~~~~~~~~~~~~~~\ \ \text{%
for~}n>2.
\end{eqnarray}
Accordingly the action on the superfield $\Phi (x,\theta )$ is computed to
\begin{eqnarray}
D_{\varepsilon }\Phi  &=&\theta \partial _{x,\varepsilon }\xi +u, \\
D_{\varepsilon }^{2}\Phi  &=&\theta \partial _{x,\varepsilon }u+\partial
_{x,\varepsilon }\xi , \\
D_{\varepsilon }^{3}\Phi  &=&\theta \partial _{x,\varepsilon }^{2}\xi
+\partial _{x,\varepsilon }u, \\
&&\vdots   \notag \\
D_{\varepsilon }^{2n-1}\Phi  &=&\theta \partial _{x,\varepsilon }^{n}\xi
+\partial _{x,\varepsilon }^{n-1}u, \\
D_{\varepsilon }^{2n}\Phi  &=&\theta \partial _{x,\varepsilon
}^{n}u+\partial _{x,\varepsilon }^{n}\xi .
\end{eqnarray}
This means for $n>2$ the derivatives acting on the fermionic as well as the
ones acting on the bosonic field are deformed. However, in general we would
like to take $\varepsilon $ to be an integer and since $\partial
_{x,\varepsilon }^{n}\xi =-i(i\xi _{x})^{\varepsilon }=0$ for $\varepsilon
=2,3,\ldots $ this does not appear to be an interesting choice.

\subsubsection{PT-symmetric superderivatives of fermionic type}

Alternatively we may define 
\begin{equation}
\hat{D}_{\varepsilon }^{n}:=D^{n-1}D_{\varepsilon }~~~~~~~~~~~~~~\ \ \text{%
for~}n>1.
\end{equation}
in which case the action on the superfield $\Phi (x,\theta )$ gives 
\begin{eqnarray}
\hat{D}_{\varepsilon }\Phi &=&\theta \partial _{x,\varepsilon }\xi +u, \\
\hat{D}_{\varepsilon }^{2}\Phi &=&\theta u_{x}+\partial _{x,\varepsilon }\xi
, \\
\hat{D}_{\varepsilon }^{3}\Phi &=&\theta \partial _{x,\varepsilon }^{2}\xi
+u_{x}, \\
&&\vdots  \notag \\
\hat{D}_{\varepsilon }^{2n-1}\Phi &=&\theta \partial _{x,\varepsilon
}^{n}\xi +\partial _{x}^{n-1}u, \\
\hat{D}_{\varepsilon }^{2n}\Phi &=&\theta \partial _{x}^{n}u+\partial
_{x,\varepsilon }^{n}\xi .
\end{eqnarray}
Thus with this choice only the terms involving the derivatives acting on
fermionic fields are $\mathcal{PT}$-symmetrically deformed, which for the
reasons mentioned at the end of the last subsection is even less exciting. \ 

\subsubsection{PT-symmetric superderivatives of bosonic type}

It is clear from the above discussion that the most interesting definitions
will be those just involving deformations of derivatives acting on the
bosonic fields. We may achieve this by defining 
\begin{eqnarray}
\tilde{D}_{\varepsilon }^{2} &:&=D_{\varepsilon }D,  \label{de1} \\
\tilde{D}_{\varepsilon }^{n} &:&=D^{n-2}D_{\varepsilon }^{2}~~~~~~~~~~~~~~\
\ \text{for~}n>2.  \label{de2}
\end{eqnarray}
In this case the action on the superfield $\Phi (x,\theta )$ turns out to be 
\begin{eqnarray}
\tilde{D}_{\varepsilon }\Phi &=&\theta \xi _{x}+u, \\
\tilde{D}_{\varepsilon }^{2}\Phi &=&\theta \partial _{x,\varepsilon }u+\xi
_{x}, \\
\tilde{D}_{\varepsilon }^{3}\Phi &=&\theta \xi _{xx}+\partial
_{x,\varepsilon }u, \\
&&\vdots  \notag \\
\tilde{D}_{\varepsilon }^{2n-1}\Phi &=&\theta \partial _{x}^{n}\xi +\partial
_{x,\varepsilon }^{n-1}u, \\
\tilde{D}_{\varepsilon }^{2n}\Phi &=&\theta \partial _{x,\varepsilon
}^{n}u+\partial _{x}^{n}\xi .
\end{eqnarray}
Thus with this choice we have achieved that only the terms involving the
derivatives acting on the bosonic fields are $\mathcal{PT}$-symmetrically
deformed.

According to the principle that any function which transforms as $\mathcal{PT%
}$: $f\rightarrow -f$ \ should be deformed as $f\rightarrow
-i(if)^{\varepsilon }$, we may also try to deform the superderivatives
directly instead of focussing on the part of it involving the ordinary
derivatives. Observing that $\mathcal{PT}$: $D\Phi \rightarrow D\Phi $, this
form of deformation can not be applied to the superderivative of first
order. However, we may apply it to higher orders. We have $\mathcal{PT}$: $%
D^{2}\Phi \rightarrow -iD^{2}\Phi $, $D^{3}\Phi \rightarrow -D^{3}\Phi $ and
therefore we may consistently define
\begin{eqnarray}
\check{D}_{\varepsilon }^{n} &:&=D^{n}\qquad \qquad \qquad \qquad \qquad
\qquad \qquad \qquad \qquad \text{for~}n=1,2 \\
\check{D}_{\varepsilon }^{3}\Phi &:&=-i(iD^{3}\Phi )^{\varepsilon }=\partial
_{x,\varepsilon }u+i\theta \varepsilon \partial _{x,\varepsilon -1}u\xi
_{xx}, \\
\check{D}_{\varepsilon }^{n} &:&=D^{n-3}\check{D}_{\varepsilon }^{3}\qquad
\qquad \qquad \qquad \qquad \qquad \qquad \qquad \text{for~}n>3.  \label{d33}
\end{eqnarray}

Taking only $\mathcal{PT}$-symmetry as a guiding principle there are of
course more possibilities. For instance, we could have also nested the
derivatives as $D_{\varepsilon }(D_{\varepsilon }\ldots D_{\varepsilon
}f)\ldots ))$, or $D_{\varepsilon }(D_{\varepsilon }\ldots \check{D}%
_{\varepsilon }^{3}f)\ldots ))$, etc. For similar reasons as stated for the
ordinary derivatives we restrain here from these choices. Alternatively we
could keep the ordinary superderivatives up to some higher order derivative,
since $D^{4n-1}\Phi \rightarrow -D^{4n-1}\Phi $ for $n\in \mathbb{N}$, but
the models we are concerned with here do not involve such high order.

We may now use either of these possibilities in any of the terms in (\ref{1}%
), giving rise to many different options to formulate $\mathcal{PT}$%
-symmetric extensions.

\subsection{Construction of new models}

We can replace the superderivatives by their deformed versions in various
different terms and in addition we may introduce different deformation
parameters in the higher order derivatives. In order to explore some of
these possibilities, let us first rewrite equation (\ref{1}) as
\begin{equation}
\Phi _{t}=-D^{6}\Phi +6D\Phi D^{2}\Phi +\lambda \Phi D^{3}\Phi -\lambda
D\Phi D^{2}\Phi ,  \label{kdvn}
\end{equation}
by using the identities $D^{2}(\Phi D\Phi )=D^{2}\Phi D\Phi +\Phi D^{3}\Phi $
and $D^{2}\Phi D\Phi =D\Phi D^{2}\Phi $. Note that these identities no
longer hold in the deformed cases, such that we would have a different
starting point when deforming (\ref{1}) directly. As discussed above, the
purely bosonic deformation is the most interesting one and we may therefore
consider
\begin{equation}
\Phi _{t}=-\tilde{D}_{\varepsilon }^{6}\Phi +6\tilde{D}_{\kappa }\Phi \tilde{%
D}_{\kappa }^{2}\Phi +\lambda \Phi \tilde{D}_{\mu }^{3}\Phi -\lambda \tilde{D%
}_{\nu }\Phi \tilde{D}_{\nu }^{2}\Phi .  \label{zs}
\end{equation}
In order to remain as generic as possible we have introduced four different
deformation parameters $\varepsilon ,\kappa ,\mu $ and $\nu $. The component
version of (\ref{zs}) reads
\begin{eqnarray}
u_{t} &=&-\partial _{x,\varepsilon }^{3}u+6u\partial _{x,\kappa }u-\lambda
\xi \xi _{xx}+\lambda u\left( \partial _{x,\mu }u-\partial _{x,\nu }u\right)
,\text{\qquad }  \label{def1} \\
\xi _{t} &=&-\xi _{xxx}+6u\xi _{x}+\lambda (\xi \partial _{x,\mu }u-u\xi
_{x}).  \label{def2}
\end{eqnarray}
The case $\mu =\nu $, $\kappa =1$ constitutes a fermionic extension of the $%
\mathcal{PT}$-symmetric deformation of the KdV-equation introduced in \cite
{AFKdV}, which is obtained for $\xi \rightarrow 0$. In turn the case $\mu
=\nu $, $\varepsilon =1$ reduces for $\xi \rightarrow 0$ to the $\mathcal{PT}
$-symmetric deformation of the KdV-equation introduced in \cite{BBCF}.
Noting how a deformed derivative transforms under a supersymmetry
transformation 
\begin{eqnarray}
\mathcal{SUSY} &:&\partial _{x,\varepsilon }u\rightarrow \partial
_{x,\varepsilon }u+i\eta \varepsilon \partial _{x,\varepsilon -1}u\xi
_{xx},\quad \\
&&\partial _{x,\varepsilon }^{3}u\rightarrow \partial _{x,\varepsilon
}^{3}u+i\eta \varepsilon (\partial _{x,\varepsilon -1}^{3}u\xi
_{xx}+2\partial _{x,\varepsilon -1}^{2}u\xi _{xxx}+\partial _{x,\varepsilon
-1}u\xi _{xxxx}),
\end{eqnarray}
it is easily seen that the equations (\ref{def1}) and (\ref{def2}) are only
invariant under the supersymmetry transformations (\ref{SUSY}) in the case $%
\mu =\nu =\kappa =\varepsilon =1$.

Instead of employing $D\rightarrow \tilde{D}_{\varepsilon }$ let us now use
the deformation $D\rightarrow \check{D}_{\varepsilon }$. An interesting
possibility is to deform just the first term in (\ref{kdvn}) and consider 
\begin{equation}
\Phi _{t}=-\check{D}_{\varepsilon }^{6}\Phi +6D\Phi D^{2}\Phi +\lambda \Phi
D^{3}\Phi -\lambda D\Phi D^{2}\Phi .  \label{xx}
\end{equation}
Using (\ref{d33}), we find $\check{D}_{\varepsilon }^{6}\Phi =\theta
\partial _{x,\varepsilon }^{3}u+i\varepsilon (\partial _{x,\varepsilon
-1}^{2}u\xi _{xx}+\partial _{x,\varepsilon -1}u\xi _{xxx})$, such that the
component version of (\ref{xx}) reads
\begin{eqnarray}
u_{t} &=&-\partial _{x,\varepsilon }^{3}u+6uu_{x}-\lambda \xi \xi _{xx},%
\text{\qquad }  \label{1q} \\
\xi _{t} &=&-i\varepsilon (\partial _{x,\varepsilon -1}^{2}u\xi
_{xx}+\partial _{x,\varepsilon -1}u\xi _{xxx})+(6-\lambda )u\xi _{x}+\lambda
\xi u_{x}.  \label{2q}
\end{eqnarray}
Thus equation (\ref{xx}) may also be viewed as yet another fermionic
extension of the $\mathcal{PT}$-symmetric deformation of the KdV-equation of 
\cite{AFKdV}, to which (\ref{1q}) reduces in the limits $\xi \rightarrow 0$
or $\lambda \rightarrow 0$. Interestingly this system is partially
supersymmetric. We find that (\ref{1q}) remains invariant under the
supersymmetry transformation (\ref{SUSY}), but (\ref{2q}) does not respect
it.

Further interesting options are of course combinations of the above, such
for instance 
\begin{equation}
\Phi _{t}=-\check{D}_{\varepsilon }^{6}\Phi +6\tilde{D}_{\kappa }\Phi \tilde{%
D}_{\kappa }^{2}\Phi +\lambda \Phi \check{D}_{\mu }^{3}\Phi -\lambda \tilde{D%
}_{\nu }\Phi \tilde{D}_{\nu }^{2}\Phi
\end{equation}
or to add $\mathcal{PT}$-invariant terms which vanish in the limit $%
\varepsilon \rightarrow 1$. We will make use of the last possibility in
order to restore full supersymmetry.

A few comments are in order: There are of course various other options, as
for instance to deform only one of the last two terms in (\ref{kdvn}),
possibly together with the first term. This would lead to a rather strange
extension, which does not reduce to any of the known $\mathcal{PT}$-extended
KdV-equations for $\xi \rightarrow 0$. These cases involve an additional
term resulting from the fact the original sKdV-equation was constructed as a
one-parameter family taking into account that the term $6uu_{x}$ can be
supersymmetrised in various alternative ways. Further options are to use the
derivatives $\hat{D}_{\varepsilon }$ or $D_{\varepsilon }$, which yield
similar equations as above with the difference that also the derivatives
acting on the $\xi $-fields are deformed, which is, however, less
interesting for the reasons mentioned above.

\section{$\mathcal{PT}$ and supersymmetric non-Hermitian Hamiltonian
deformations}

Let us now recall the original motivation to consider $\mathcal{PT}$%
-symmetrically extended models, which was to exploit the feature that
unbroken $\mathcal{PT}$-symmetry guarantees the reality of the corresponding
spectrum. In this spirit it is highly desirable to discriminate between the
models, which are Hamiltonian systems and those which are not. It is well
known that the sKdV-equation admits a Hamiltonian description for $\lambda
=2 $, see \cite{Mat}, and it is interesting to investigate whether this
feature survives the deformation.

Making use of the usual properties for the Berezin integral $\int d\theta =0$%
, $\int d\theta \theta =1$, we consider the Hamiltonian 
\begin{eqnarray}
H_{\varepsilon } &=&\int d\mu \left[ \Phi (D\Phi )^{2}+\frac{1}{%
1+\varepsilon }D^{2}\Phi \check{D}_{\varepsilon }^{3}\Phi \right]
\label{hdef} \\
&=&\int dx\left[ u^{3}-2\xi \xi _{x}u-\frac{1}{1+\varepsilon }%
(iu_{x})^{\varepsilon +1}-\frac{\varepsilon }{1+\varepsilon }%
(iu_{x})^{\varepsilon -1}\xi _{x}\xi _{xx}\right] ,  \label{zsc}
\end{eqnarray}
where we abbreviated $\int dxd\theta =:\int d\mu $. This Hamiltonian is a
deformed version of the sKdV Hamiltonian \cite{Mat} and in addition a
supersymmetrised version of the $\mathcal{PT}$-symmetrically deformed
Hamiltonian \cite{AFKdV}, as it reduces to these Hamiltonians in the limits $%
\varepsilon \rightarrow 1$ and $\xi \rightarrow 0$, respectively. By
construction $H_{\varepsilon }$ is $\mathcal{PT}$-symmetric, but in addition
it is also supersymmetic, which is most easily verified for the component
version (\ref{zsc})
\begin{equation}
\mathcal{SUSY}:H_{\varepsilon }\rightarrow H_{\varepsilon }+\eta \int
dx\partial _{x}\left( \xi u^{2}+\frac{i^{\varepsilon -1}}{1+\varepsilon }%
u_{x}^{\varepsilon }\xi _{x}\right) =H_{\varepsilon }.
\end{equation}
This means we can also think of $H_{\varepsilon }$ as a new supersymmetric
version of the KdV-Hamiltonian.

Unlike as for the KdV-equation, which admits a bi-Hamiltonian structure \cite
{Magri}, see also \cite{Das}, the sKdV-equation is known to possess only one
such structure \cite{Mat}, which respects supersymmetry. The Poisson
brackets are defined as
\begin{equation}
\left\{ F(\mu ),G(\mu ^{\prime })\right\} :=\int d\mu _{0}\frac{\delta F(\mu
)}{\delta \Phi (\mu _{0})}D_{\mu _{0}}\frac{\delta G(\mu ^{\prime })}{\delta
\Phi (\mu _{0})}.  \label{vv}
\end{equation}
Using the same Poisson bracket structure gives rise to a deformed equation
of motion. With definition (\ref{vv}) we may then compute the corresponding
flow as 
\begin{eqnarray}
\Phi _{t} &=&\left\{ \Phi (\mu ),H\right\} =D\frac{\delta H}{\delta \Phi }=D%
\left[ \frac{\delta \int d\mu \mathcal{H}}{\delta \Phi }\right] , \\
&=&D\frac{\partial \mathcal{H}}{\partial \Phi }+D^{2}\frac{\partial \mathcal{%
H}}{\partial (D\Phi )}-D^{3}\frac{\partial \mathcal{H}}{\partial (D^{2}\Phi )%
}-D^{4}\frac{\partial \mathcal{H}}{\partial (D^{3}\Phi )}+\ldots
\end{eqnarray}
For the Hamiltonian (\ref{hdef}) we find
\begin{equation}
\Phi _{t}=4D\Phi D^{2}\Phi +2\Phi D^{3}\Phi -\frac{1}{1+\varepsilon }\left[ 
\check{D}_{\varepsilon }^{6}\Phi +i\varepsilon D^{4}(D^{2}\Phi \check{D}%
_{\varepsilon -1}^{3}\Phi )\right] ,  \label{b1}
\end{equation}
with corresponding component version 
\begin{eqnarray}
u_{t} &=&6uu_{x}-\partial _{x,\varepsilon }^{3}u-2\xi \xi _{xx}+\frac{%
\varepsilon -\varepsilon ^{2}}{1+\varepsilon }\left[ \partial
_{x,\varepsilon -2}^{3}u\xi _{x}\xi _{xx}+\partial _{x,\varepsilon
-2}^{2}u\xi _{x}\xi _{xxx}+\partial _{x}(\partial _{x,\varepsilon -2}u\xi
_{x}\xi _{xxx})\right] ,  \notag  \label{b2} \\
\xi _{t} &=&4u\xi _{x}+2\xi u_{x}-\frac{i\varepsilon }{1+\varepsilon }\left(
3\partial _{x,\varepsilon -1}^{2}u\xi _{xx}+2\partial _{x,\varepsilon
-1}u\xi _{xxx}+\partial _{x,\varepsilon -1}^{3}u\xi _{x}\right) .  \label{b3}
\end{eqnarray}
As we expect (\ref{b1}) and (\ref{b2}), (\ref{b3}) reduce to (\ref{1}) and (%
\ref{2}), (\ref{3}), in the limit $\varepsilon \rightarrow 1$, respectively.

\section{Conclusion}

We have discussed various possibilities to introduce $\mathcal{PT}$%
-symmetrically deformed superderivatives. The most interesting cases are
those just involving deformed derivatives acting on the bosonic field, i.e. $%
\tilde{D}_{\varepsilon }^{n}$ and $\check{D}_{\varepsilon }^{n}$ as defined
in (\ref{de2}) and (\ref{d33}), respectively. We have demonstrated that
these derivatives can be employed very systematically to construct new $%
\mathcal{PT}$-symmetric extensions of the sKdV-equation. Most of these
extensions are mere fermionic extensions that is they involve fermionic
superfields, but do not preserve the invariance under a supersymmetry
transformation. Remarkably it is also possible to find genuinely
supersymmetric extensions. Furthermore, these models allow for a Hamiltonian
formulation. This means we may also think of this latter models as new
supersymmetrized versions of the KdV-equation.

Clearly with regard to these new models there are many interesting questions
left to be explored. It remains to be settled whether these models possess
non-trivial higher charges and if the conservation laws survive the
deformation procedure \cite{Mat,BLR}. Possibly the new models are even
integrable. Nonetheless, even when they turn out to be non-integrable one
may exploit the rich properties of the underlying integrable model and treat
the new models as perturbations of the former. This is somewhat similar in
spirit as studying non-integrable quantum field theories as perturbations of
integrable models, see e.g. \cite{Giu}. Further interesting properties to
investigate are the nature of the solutions these equations possess, what
type of additional symmetries they allow \cite{BBD} etc.

Besides these issues centered around the sKdV equation one may of course use
the deformed superderivatives in other contexts to construct new $\mathcal{PT%
}$-symmetric deformations in the same spirit. Most immediate would be to
consider the sKdV-equation involving bosonic rather than fermionic
superfields and its N=2 version.

\end{document}